\documentclass[pra,twocolumn,amsmath,amssymb,superscriptaddress]{revtex4-1}
\usepackage{epsfig, graphicx,graphics,amsmath,amssymb,float}
\usepackage[T1]{fontenc}
\usepackage[latin9]{inputenc}
\usepackage{amsmath}
\usepackage{amssymb}
\usepackage{xcolor}
\usepackage{amscd}
\usepackage{bm}
\usepackage{bbold}
\usepackage{psfrag}

\usepackage{bbm}

\usepackage[caption=false]{subfig}
\usepackage{subfig}
\usepackage{color}
\usepackage[bookmarks=true,colorlinks,linkcolor=blue,urlcolor=blue,citecolor=blue]{hyperref}

\DeclareMathOperator{\sgn}{sgn} %definir a funcao sinal. Serve pra definir qlq funcao que nao esta pre-definida

\newcommand{\be}{\begin{equation}}
\newcommand{\ee}{\end{equation}}
\newcommand{\bea}{\begin{eqnarray}}
\newcommand{\eea}{\end{eqnarray}}

\newcommand{\mc}{\mathcal}
\newcommand{\mb}{\mathbf}

\begin{document}

%\title{Dynamics of the Kondo effect after connecting a magnetic impurity to interacting metallic chains}
%\title{Quench dynamics of the Kondo effect  in a Luttinger liquid}
\title{Quench dynamics and relaxation of a spin coupled to interacting leads }

\author{Helena Bragan\c{c}a}
\affiliation{Instituto de F\'{i}sica and International Center for Physics, Universidade de Bras\'{i}lia, Bras\'{i}lia 70919-970, DF, Brazil}
\affiliation{Departamento de F\'isica, Universidade Federal de Minas Gerais, C. P. 702, 30123-970, Belo Horizonte, MG, Brazil}
\author{M. F. Cavalcante}
\affiliation{Departamento de F\'isica, Universidade Federal de Minas Gerais, C. P. 702, 30123-970, Belo Horizonte, MG, Brazil}
\author{R. G. Pereira}
\affiliation{International Institute of Physics and Departamento de F\'isica Te\'orica e Experimental, Universidade Federal do Rio Grande do Norte, 59072-970 Natal-RN, Brazil}
\author{Maria C. O. Aguiar}
\affiliation{Departamento de F\'isica, Universidade Federal de Minas Gerais, C. P. 702, 30123-970, Belo Horizonte, MG, Brazil}

\date{\today}

\begin{abstract}
We study a   quantum quench in which a magnetic impurity is suddenly coupled to Hubbard chains, whose low-energy physics is described by  Tomonaga-Luttinger liquid theory. Using the time-dependent density-matrix renormalization-group (tDMRG) technique, we analyze the  propagation   of charge, spin and entanglement in the chains after the quench  and relate the light-cone velocities   to the dispersion of holons and spinons. We find that the local magnetization at the impurity site decays faster if we increase the interaction in the chains, even though the spin velocity decreases. We derive an analytical expression for the relaxation of the impurity magnetization  which is in good agreement with the tDMRG results at intermediate timescales, providing valuable insight into the time evolution  of the Kondo screening cloud in interacting systems. 

\end{abstract}

\pacs{pacs}

\maketitle

\section{Introduction}

The interaction between a magnetic impurity and a non-magnetic metallic host %, which leads to anomalous resistivity behavior, 
 is one of the fundamental problems in many-body physics~\cite{Kondobook}. 
A deep understanding of the Kondo effect has been  achieved thanks to the numerical  renormalization group~\cite{NRG},  the exact Bethe ansatz solution~\cite{BA}, and conformal field theory techniques~\cite{CFT}. A hallmark of the Kondo effect is the emergence of a characteristic  scale, the Kondo temperature $T_K$, which varies exponentially with the exchange coupling between the impurity and a Fermi liquid metal. At   temperatures below $T_K$, perturbation theory in the exchange coupling breaks down, and the properties of the system are governed by the formation of a   singlet state between the   impurity and the conduction electrons \cite{Nozieres1974}.   Kondo physics is also manifested in the spatial dependence of spin correlations \cite{Eq87,Eq06,Barzykin1996,Eq07,Holzner2009,affleck2010kondo}, which change qualitatively over  distances of the order of  $\xi_K=\hbar v_F/(k_BT_K)$, where $v_F$ is the Fermi velocity of the conduction electrons and $\hbar$ and $k_B$ are the Planck and Boltzmann constants, respectively. The length scale $\xi_K$ is interpreted  as the size of  the Kondo screening cloud, whose  effects  have been observed  recently in a mesoscopic device \cite{Borzenets2020}.

Experiments with  quantum dots coupled to electron reservoirs \cite{Eq09,dot2}  opened the way for controllable realizations and  inspired studies of the nonequilibrium dynamics   of   Kondo systems   \cite{1kondo, 2kondo, TD-NRG, 3kondo, 4kondo, Tureci2011,Costi2017, Novo, Krivenko2019}. The active research on this topic has also been boosted by recent efforts to simulate the Kondo effect with ultracold atoms \cite{ultracold2, nishida2013,riegger2018localized,Kanasz-Nagy2018}. A typical quench protocol in this context  consists of    switching on the coupling between an initially spin-polarized  impurity and a metallic lead   \cite{1kondo,Medvedyeva2013,Lechtenberg2014,Nuss2015}.  One then observes a real-time decay of the impurity magnetization accompanied by  the buildup of Kondo correlations   over a time scale $\tau_K=\hbar/(k_BT_K)$ after the quench. By analogy with the Kondo effect, 
the scaling behavior in the  time dependence as the system approaches equilibrium has also been studied  for the resonant level model~\cite{Vasseur2013,Kennes2014,Ghosh2015}. 

The usual description of the Kondo effect within the single-impurity Anderson model~\cite{Kondobook} takes into account the local interaction at the quantum dot, but neglects electron-electron interactions in the bulk. In higher dimensions, this approximation is justified by Landau's Fermi liquid theory, where the bulk degrees of freedom are associated   with weakly interacting fermionic quasiparticles. However, if the leads are interacting one-dimensional systems, as in the case of quantum wires \cite{Deshpande2010}, Fermi liquid theory must be replaced by Tomonaga-Luttinger liquid (TLL) theory~\cite{Giamarchi}. In a TLL, the elementary excitations are spin-charge-separated bosonic modes and  correlation functions decay as power laws with interaction-dependent exponents. Despite the spin-charge separation in the bulk, the Kondo effect in a TLL can still be affected by interactions in the charge sector \cite{Lee1992,Furusaki1994,Frojdh1995,Fabrizio1995,Hallberg1995,Egger1998}.

In this work we investigate the nonequilibrium dynamics after a local quench in which  a spin-polarized electron is suddenly coupled to a correlated chain described by a repulsive Hubbard model. We are particularly  interested in comparing numerical results obtained by  time-dependent density matrix renormalization group (tDMRG) methods with the predictions from an effective field theory for the coupling of a localized spin to a TLL.   We first study the problem using bosonization to describe the quench within TLL theory, following the  approach of Ref. \cite{Schiro2015}.  Applying perturbation theory in the Kondo coupling $J_K$, we derive an analytical expression for the decay of the impurity  magnetization valid   up to   intermediate times, which are greater  than the microscopic time scale $\hbar/J_K$ but shorter than the Kondo time $\tau_K$. Second, we simulate the nonequilibrium dynamics  in the lattice model   using  tDMRG.  Analyzing  the  propagation of correlations and entanglement entropy, we observe two light cones with distinct velocities, which can be identified with the maximum velocities of elementary charge and spin excitations in the Hubbard chain.  The numerical results confirm the field theory prediction of a time regime where the impurity magnetization scales logarithmically with time. On the other hand, they also reveal a strong dependence on the interaction strength which we ascribe to a renormalization of the high-energy cutoff in the field theory, rather than to the Luttinger parameter in the charge sector.

The remainder of the paper is organized as follows. In Sec. \ref{model}, we introduce the time-dependent Anderson impurity model with interacting leads and describe the quench protocol. In Sec.  \ref{sec:outline}, we use the TLL theory to derive an analytical expression for the real-time decay of the impurity magnetization from perturbation theory in the Kondo coupling. Section \ref{numresults} presents our tDMRG results. We discuss the propagation of perturbations after the quench in relation to the exact velocities of elementary excitations and analyze the relaxation of local observables at the impurity site.  Our concluding remarks can be found in Sec. \ref{sec:conclusions}. Finally, Appendix \ref{1stappendix} contains some details of the analytical calculation outlined in Sec. \ref{sec:outline}. Hereafter we set $\hbar=k_B=1$.

\section{Model and quench protocol} \label{model}

\begin{figure}[t]
  \begin{center}
 \includegraphics[width=0.95\columnwidth]{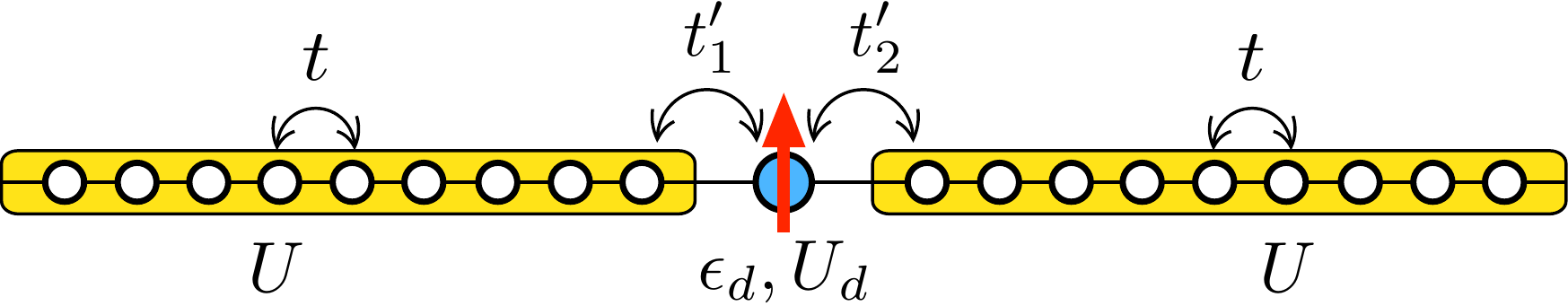}
 \end{center}
 \caption{Schematic representation of the quantum quench in the Anderson impurity model with interacting leads. At the initial time, the impurity site with energy $\epsilon_d$ and interaction $U_d$, occupied by a spin-up electron, is connected to the left and right wires via  hybridization couplings $t'_1$ and $t_2'$, respectively. The wires are described by a quarter-filled Hubbard model with hopping parameter $t$ and on-site interaction $U$. }
  \label{geometry}
  \end{figure}

We consider the  setup   shown in Fig. \ref{geometry}, described by the time-dependent Hamiltonian
\begin{equation}
  H(\tau)=H_{\rm leads}+H_{\rm imp}+\Theta(\tau)H_{\rm hyb}, \label{HamDin}
\end{equation}
where $\Theta(\tau)$ denotes the Heaviside step function. 
The first term represents the Hamiltonian for the conduction electrons in the interacting leads, $H_{\rm leads}=\sum_{\ell=1}^2H_\ell$,
 where the index $\ell=1,2$ labels the left and right wires, respectively.  Each wire is described by a Hubbard model:
\bea
H_{1}&=&-t\sum_{i=-L_1}^{-2}\sum_{ \sigma}(c^{\dagger}_{i\sigma}c^{\phantom\dagger}_{i+1\sigma}+\textrm{h.c.})+ U \sum_{i=-L_1}^{-1}n_{i\uparrow} n_{i\downarrow} , \nonumber\\
H_{2}&=&-t\sum_{i=1}^{L_2-1}\sum_{ \sigma}(c^{\dagger}_{i\sigma}c^{\phantom\dagger}_{i+1\sigma}+\textrm{h.c.})+U \sum_{i=1}^{L_2}n_{i\uparrow} n_{i\downarrow} , \label{RHam}
\eea
where $c_{i\sigma}$ annihilates an electron with spin $\sigma=\uparrow,\downarrow$ at site $i$,   $n_{i\sigma}=c^{\dagger}_{i\sigma}c^{\phantom\dagger}_{i\sigma}$ are number operators,  $t$ is the hopping parameter, and $U>0$ is the  on-site interaction strength. The number of sites is  $L_1$ for  the chain on the left and $L_2$ for the chain on the right, and we impose open boundary conditions at the chain ends. Throughout this work, we set $t=1$, which defines the unit of energy. We shall work at fixed average density $\rho=N_1/L_1=N_2/L_2$, where $N_\ell$ is the number  of electrons in each decoupled chain in the initial state. At half filling, an infinitesimal $U>0$ drives the system into a Mott insulating phase \cite{Giamarchi}. To  consider interacting metallic leads, we set the density to quarter filling, $\rho=1/2$. 

The  Hamiltonian for the impurity site, $i=0$, is given by \be
 H_{\rm imp}=\epsilon_d  n_0+U_d n_{0 \uparrow} n_{0 \downarrow},
\ee
where $n_0=n_{0\uparrow}+n_{0\downarrow}$, $\epsilon_d$ is the energy shift of the localized state with respect to the Fermi level in the leads, and $U_d$ is the local interaction. To have a local moment at the impurity site,   we consider $\epsilon_d<0$ and $U_d>0$.

Finally, the    hybridization between the impurity and the chains is described by
\begin{equation}
 H_{\rm hyb}=-t'_1\sum_\sigma c^\dagger_{-1 \sigma}c^{\phantom\dagger}_{0 \sigma}-t'_2\sum_\sigma c^\dagger_{0 \sigma}c^{\phantom\dagger}_{1\sigma}+\textrm{h.c.}, \label{HamCoup} 
\end{equation}
where $t'_1$ and $t'_2$ are the hopping parameters between   the impurity site and the nearest sites in the left or right chains, respectively.  While this model describes the general case of  asymmetric tunnelling amplitudes, we shall focus on two  special limits. First,  in the maximally asymmetric case, we set  $t_1'=0$ and $t_2'\neq0$.  The impurity site is then coupled only to the first site  of the right wire  and we can forget about  the left wire.  Second, for $t_1'=t_2'$, the impurity site is coupled symmetrically to both wires. In this case, hopping through the impurity site allows for electron transport between the wires.

We are mainly interested in the strong Coulomb blockade regime $t_{1,2}^\prime\ll -\epsilon_d,U_d$, where states in   the low-energy subspace contain a singly occupied impurity site, $n_0=1$. We then  apply a Schrieffer-Wolff transformation to integrate out high-energy processes that change the occupation of the impurity site \cite{Kondobook}. As a result, model~\eqref{HamDin} can be mapped onto the effective Hamiltonian   \be
H_{\rm eff}(\tau)=H_{\rm leads}+\Theta(\tau)H_{K},\label{Heff}
\ee
where \cite{Simon2003}\bea
H_{K}&=&J_K\mathbf S_0\cdot \left(\kappa_1c^\dagger_{-1}+\kappa_2c^\dagger_{1}\right)\frac{\boldsymbol\sigma}{2}\left(\kappa_1c^{\phantom\dagger}_{-1}+\kappa_2c^{\phantom\dagger}_{1}\right)\nonumber\\
&&+V \left(\kappa_1c^\dagger_{-1}+\kappa_2c^\dagger_{1}\right)\left(\kappa_1c^{\phantom\dagger}_{-1}+\kappa_2c^{\phantom\dagger}_{1}\right).\label{HKondo}
\eea
Here $\mathbf S_0=c^\dagger_{0}(\boldsymbol\sigma/2)c^{\phantom\dagger}_{0}$ is the spin operator for the electron at the impurity site and   $\boldsymbol\sigma$ is the vector of Pauli matrices acting on the two-component spinor $c_i=(c_{i\uparrow},c_{i\downarrow})^t$. The antiferromagnetic Kondo coupling $J_K>0$ is given by \be
J_K=2\left[(t_1')^2+(t_2')^2\right]\left(\frac1{-\epsilon_d}+\frac{1}{U_d+\epsilon_d}\right),\label{JK}
\ee
and    the amplitude $V$ of the potential scattering term is given by\be
V=\frac{(t_1')^2+(t_2')^2}{2}\left(\frac1{-\epsilon_d}-\frac{1}{U_d+\epsilon_d}\right). 
\ee
The dimensionless parameters $\kappa_{1,2}$ in Eq. (\ref{HKondo}) are \bea
\kappa_1=\frac{t_1'}{\sqrt{(t_1')^2+(t_2')^2}},\quad \kappa_2=\frac{t_2'}{\sqrt{(t_1')^2+(t_2')^2}}.
\eea
For $t'_1=0$ and $t_2'\neq0$, the impurity spin couples to the spin density at  the boundary of the right wire.  For $t_1'=t_2'$,  it couples to the symmetric orbital at sites $i=\pm1$. 
The potential scattering can be cancelled at lowest order in $t_{1,2}^\prime/U_d$  if we set $\epsilon_d=-U_d/2$. Note, however, that a nonzero $V$ may be generated at higher orders  because particle-hole symmetry is broken for any electron density away from half filling.

In the quench protocol, we assume that for times $\tau<0$ the system is prepared  in the state\be
|\Psi_0\rangle=|\textrm{GS}\rangle_1\otimes \left|\uparrow \right\rangle \otimes|\textrm{GS}\rangle_2 , 
\ee
where $|\textrm{GS}\rangle_\ell$ for $\ell=1,2$ is the ground state of the disconnected  Hubbard chains and  $\left|\uparrow \right\rangle$ is the polarized state of the impurity spin. This initial state is not an eigenstate of   Hamiltonian (\ref{HamDin}) for $\tau>0$. After we switch on the hybridization couplings,   the  state evolves nontrivially  according to \be
\left|\Psi(\tau)\right\rangle=e^{-i\tau(H_{\rm leads}+H_{\rm imp}+H_{\rm hyb})}\left|\Psi_0\right\rangle.
\ee
Since this is a local quench, we expect that,  in the thermodynamic limit and for sufficiently long times, local observables will relax to   the corresponding expectation values in the ground state of the post-quench Hamiltonian. The ground state of $H(\tau>0)$ in Eq. (\ref{HamDin})   at quarter filling was studied numerically in Ref.~\cite{equilibrium}. For  noninteracting chains,  $U=0$, one finds that the Kondo regime, where $\langle n_0\rangle\approx 1$ and $\langle (S_0^z)^2\rangle\approx1/4$, is reached with good approximation for $t'=0.5$ and $U_d\gtrsim 5$. Smaller values of $U_d$ lead to a mixed valence regime with significant deviations from   single occupancy in the impurity site. Importantly, for a fixed  value of $U_d$,  increasing the  interaction $U>0$ in the chains suppresses charge fluctuations and drives the system closer to the Kondo regime \cite{equilibrium}.

\section{Field theory approach} \label{sec:outline}

Deep in the Kondo regime, we can use the  Hamiltonian in Eq.  (\ref{Heff}) to study the time evolution of observables which do not entail  charge fluctuations at the impurity site.  In particular, we shall be interested in the time-dependent impurity magnetization\be
m_0(\tau)=\langle \Psi(\tau)|S_0^z|\Psi(\tau)\rangle.\label{m0tau}
\ee
The latter must decay to zero for $\tau\to\infty$, signalling the formation of the Kondo screening cloud. In this Section, we calculate the time dependence of $m_0(\tau)$ using the low-energy effective field theory and perturbation theory in the Kondo coupling. Our calculation for  electrons with spin  is inspired by the approach  of Ref. \cite{Schiro2015} for spinless fermions.

\subsection{Effective Hamiltonian}

We start with the Luttinger model for the disconnected interacting leads. Here we consider two semi-infinite chains, corresponding to the limit $L_{1,2}\to\infty$. To describe the low-energy modes in the wires, we take the continuum limit and expand the fermionic field operators in terms of right (R) and left (L) movers \cite{Giamarchi}\be
c_{j\sigma}\sim c_{\ell,\sigma}(x)=e^{ik_{F}x}\psi_{R,\ell,\sigma}(x) + e^{-ik_{F}x}\psi_{L,\ell,\sigma}(x),
\ee
where $c_{1,\sigma}(x)$ is defined for $x<0$  and $c_{2,\sigma}(x)$  for $x>0$. At quarter filling,   the Fermi momentum is $k_F=\pi/4$, where we set the lattice spacing  to unity.  The open boundary condition, $c_{\ell,\sigma}(0)=0$, can be cast as a constraint on   the chiral fermionic modes  in each wire  \cite{Fabrizio1995}
\be
\psi_{L,\ell,\sigma}(x)=-\psi_{R,\ell,\sigma}(-x).
\ee
Thus, for instance, the left  mover in  wire $\ell=2$ can be regarded as the analytic continuation of the right mover to the negative-$x$ axis. Likewise, we may choose the right mover in wire $\ell=1$ to be the analytic continuation of the corresponding left mover to the positive-$x$ axis. These relations allow us to work with a single chiral field in each wire redefined in the domain $x\in \mathbbm R$. It will be convenient to use the two-component  spinors \bea
\psi^\dagger_{1}(x)&=&(\psi^\dagger_{L,1,\uparrow}(-x),\psi^\dagger_{L,1,\downarrow}(-x)),\nonumber\\
\psi^\dagger_{2}(x)&=&(\psi^\dagger_{R,2,\uparrow}(x),\psi^\dagger_{R,2,\downarrow}(x)).
\eea

In the continuum limit, the Hamiltonian for the interacting leads becomes the Luttinger model with an open boundary  \cite{Fabrizio1995}, $H_{\ell}\approx H_{\ell }^{\rm LL}$, with \bea
H_\ell^{\rm LL}&=&v_F\int_{-\infty}^{\infty} dx\,\left[\psi^\dagger_{\ell}(-i\partial_x)\psi^{\phantom\dagger}_{\ell}\right.\nonumber\\
&&\left.+\frac{g}2\rho_{\ell}(x)\rho_{\ell}(x)+\frac{g}2 \rho_{\ell }(x)\rho_{\ell}(-x)\right],
\eea
where $v_F=2\sin k_F$ is the Fermi velocity of the noninteracting system, $g=U/ v_F$  is a dimensionless parameter, and $\rho_{\ell}=\psi^\dagger_{\ell}\psi^{\phantom\dagger}_{\ell}$ is a density operator.

Using Abelian bosonization~\cite{Giamarchi,gogolin}, we write the fermionic field    in terms of charge and spin boson fields:
\begin{equation}
\psi_{\ell}(x)\sim\frac{1}{\sqrt{2\pi\alpha}}\left(\begin{array}{c}e^{-i\sqrt{\frac{\pi}{2}}[\phi_{\ell,c}(x)+\phi_{\ell,s}(x)]}\\
e^{-i\sqrt{\frac{\pi}{2}}[\phi_{\ell,c}(x)-\phi_{\ell,s}(x)]} \end{array} \right),    
\end{equation}
where $\alpha$ is a short-distance cutoff and $\phi_{\ell,c/s}={(\phi_{\ell,\uparrow}\pm \phi_{\ell,\downarrow})/\sqrt{2}}$
obey the commutation relations
\begin{equation}
[\phi_{\ell,\lambda}(x),\phi_{\ell',\lambda'}(y)]=i\delta_{\ell\ell'}\delta_{\lambda\lambda'}\sgn(x-y).    
\end{equation}
In terms of bosonic annihilation operators $\eta_{\ell,\lambda, q}$ with momentum $q>0$, the fields $\phi_{\ell,\lambda}(x)$ are given by 
\begin{equation}
\phi_{\ell,\lambda}(x)=\frac{1}{\sqrt{L}}\sum_{q>0}\frac{e^{-\frac{\alpha }{2}q}}{\sqrt{q}}[ z_{\lambda q}(x)\eta_{\ell,\lambda, q} + z^{*}_{\lambda q}(x)\eta^{\dagger}_{\ell,\lambda, q}],    \label{modeexp}
\end{equation}
where $z_{\lambda q}(x) = (1/\sqrt{K_{\lambda}}) \cos(qx) + i\sqrt{K_{\lambda}}\sin(qx)$, with $K_\lambda$ the Luttinger parameter in the charge or spin sector for $\lambda=c,s$, respectively. The SU(2)  spin-rotation   symmetry of the Hamiltonian for the leads  fixes $K_s=1$ \cite{Giamarchi,gogolin}. For the charge sector, bosonization gives $K_c\approx 1-g/\pi$ to first order in $g\ll1$. The exact value of the charge Luttinger parameter can be obtained from the Bethe ansatz solution of the Hubbard model \cite{LiebWu,essler2005}. For repulsive interactions, one finds $K_c<1$, with  $K_c=1$ for $U=0$ and $K_c\to1/2$ for $U\to\infty$. 
Using the mode expansion in Eq. (\ref{modeexp}), we   diagonalize   the Luttinger Hamiltonian in  the form
\begin{equation}
H_{\ell}^{\rm LL}=\sum_{\lambda=c,s}\sum_{q>0}v_{\lambda}q\eta^{\dagger}_{\ell,\lambda, q}\eta^{\phantom\dagger}_{\ell,\lambda, q},   \label{diagonal}
\end{equation}
with $v_{c/s}$ being the velocities of the charge   and spin  bosonic modes. Like the Luttinger parameters, the exact velocities can be determined using the Bethe ansatz solution. In particular, in the strong coupling limit $U\to\infty$,   the ground-state wavefunction can be expressed as a product of a Slater determinant of spinless fermions and the spin wavefunction of the  spin-$1/2$ Heisenberg chain~\cite{Ogata1990}. In this limit,  the spin velocity vanishes,  $v_s\sim t^2/U\to0$, while the charge velocity approaches $v_c=2\sin k_F$ \cite{Ogata1990,Schulz1991}.

The Kondo coupling involves the fermionic fields at sites $i=\pm1$. Using the boundary condition, we get\bea
c_{1\sigma}&\sim& e^{ik_F}\psi_{R,2,\sigma}(0)+e^{-ik_F}\psi_{L,2,\sigma}(0)\nonumber\\
&=&2i\sin k_F\,\psi_{R,2,\sigma}(0).
\eea
Likewise, $c_{-1\sigma}\sim 2i\sin k_F\,\psi_{L,1,\sigma}(0)$. We can now write the Kondo coupling in the weak coupling limit $J_K\ll t$. We will distinguish between the two cases of interest. First, for $t_1'=0$ we obtain
\be
H^{(1)}_{K}=\lambda_{K}\mathbf S_{0}\cdot  :\psi_{2}^{\dagger}(0)\frac{\bm{\sigma}}{2}\psi_{2}^{\phantom\dagger}(0):,   \label{H1K}
\ee
where $:...:$ denotes normal ordering and we have used $k_F=\pi/4$ and $\epsilon_d=-U_d/2$ to obtain the bare Kondo coupling $ \lambda_K=4J_K\sin^2k_F=16(t_2')^2/U_d$ to lowest order in $t_2'/U_d$. Second, for $t_1'=t_2'$,  we obtain
\be
H^{(2)}_{K}=  \lambda_{K}\mathbf S_{0}\cdot \sum_{\ell,\ell'} :\psi_{\ell}^{\dagger}(0)\frac{\bm{\sigma}}{2}\psi_{\ell'}^{\phantom\dagger}(0):,   \label{JK1}
\ee
with the same bare $ \lambda_K$  as  in Eq. (\ref{H1K}). Note that the terms with $\ell\neq\ell'$ in $H^{(2)}_{K}$ account for  processes which transfer electrons between the wires. 

In  bosonized form, the Kondo Hamiltonian for the impurity coupled to a single wire becomes 
\bea
H^{(1)}_{K}&=&\frac{\lambda_{K}}{2}\left[{S}^{+}_{0}\frac{e^{-i\sqrt{2\pi}\phi_{2s}(0)}}{2\pi\alpha}+ {S}^{-}_{0}\frac{e^{i\sqrt{2\pi}\phi_{2s}(0)}}{2\pi\alpha}\nonumber\right.\\
&&\left. +  {S}^{z}_{0}\frac{1}{\sqrt{2\pi}}\partial_x\phi_{2s}(0)\right].\label{HK2}
\eea
In this case, the Kondo interaction   does not   involve the charge boson.  In fact, spin-charge separation is preserved  when the impurity spin is coupled to the open boundary of a TLL \cite{Fabrizio1995,Pereira2008}. As a consequence, the equilibrium Kondo effect in this geometry does not exhibit anomalous scaling associated with the Luttinger parameter $K_c<1$ in the charge sector. This is in contrast with the Kondo effect in a TLL studied in Refs. \cite{Lee1992,Furusaki1994}, where the impurity spin is coupled to the electron spin density  in the middle of an infinite wire. However, the Kondo effect described by Eq. (\ref{HK2}) can still be affected by interactions in the bulk as the latter can renormalize the bare Kondo coupling $\lambda_K$, or equivalently the nonuniversal constant $\alpha$ in prefactor of the boundary spin operators. We shall see that is indeed the case when we analyze the numerical results in Sec. \ref{numresults}.

When the impurity spin is symmetrically coupled  to both wires, we obtain   
\begin{equation}
H^{(2)}_{K}=\frac{\lambda_{K}}{2}({S}^{+}_{0}F + {S}^{-}_{0}F^{\dagger} +  {S}^{z}_{0}G),
\label{hk}
\end{equation}
where the boundary operators $F$ and $G$ are given by
\bea
F&=&\frac{1}{2\pi\alpha } \left[e^{-i\sqrt{2\pi}\phi_{1s}(0)}+e^{-i\sqrt{2\pi}\phi_{2s}(0)}\right]\nonumber\\
&&+\mc C\frac{e^{-i\sqrt{\pi}\phi_{s}^+(0)}}{\pi \alpha}\cos\left[\sqrt{\pi}\phi_{c}^-(0)\right],\label{Fx}\\
G&=&\sqrt{\frac{1}{\pi}}\partial_x\phi_{s}^+(0)\nonumber\\
&&-\frac{2\mc C}{\pi \alpha}\sin\left[\sqrt{\pi}\phi_{c}^-(0)\right]\sin\left[\sqrt{\pi}\phi_{s}^-(0)\right].\label{Gx}
\eea
Here $\phi_\lambda^\pm=(\phi_{1\lambda}\pm\phi_{2\lambda})/\sqrt2$ are symmetric and antisymmetric combinations with respect to exchanging the wires. In addition to the spin-only terms analogous to Eq. (\ref{HK2}), the Kondo interaction for this geometry   contains operators that involve    the charge bosons and  are associated with tunnelling between the wires.  Here we have introduced a nonuniversal constant $\mc C$ because  the tunnelling terms renormalize differently than the ones that scatter electrons back into the same wire  \cite{Fabrizio1995}. Given that the Kondo interaction involves the charge boson, in this case we should expect some explicit interaction dependence as the Luttinger parameter $K_c$ must show up in the exponent of correlators for the boundary operators.

  \subsection{Decay of the impurity magnetization after the quantum quench}

We now turn to the calculation of  $m_0(\tau)$.  In the effective field theory, the ground state of the disconnected leads is a vacuum of the bosonic modes of the TLLs. Thus, the initial state is written as $\vert\Psi_{0} \rangle=\vert 0\rangle_{1}\otimes\left\vert\uparrow\right\rangle\otimes \vert 0\rangle_{2}$, where $\vert 0\rangle_{\ell}$ for $\ell=1,2$ obey $\eta_{\ell,\lambda,q}\vert 0\rangle_{\ell}=0$ for both spin and charge modes ($\lambda=c,s$) and any $q>0$. 

Let us first consider the case in which the impurity is coupled to a single wire.  The time evolution   for $\tau>0$ is governed by     the effective   Hamiltonian  $H_0+H_{K}^{(1)}$, with $H_0=H_{2}^{\rm LL}$. In the interaction picture, we write \be
m_0(\tau)=\langle \Psi_I(\tau)|S^z_{0,I}(\tau)|\Psi_I(\tau)\rangle. 
\ee
Here $\hat{O}_I(\tau)=e^{i H_0 \tau}\hat{O}e^{-i H_0 \tau}$ denotes an operator evolved with the unperturbed Hamiltonian, while  the  state evolves with the Kondo interaction in the form
\begin{equation}
\vert\Psi_I(\tau)\rangle = T\exp\left[-i\int_{0}^{\tau}d\tau'{H}^{(1)}_{K,I}(\tau')\right]\vert\Psi_{0}\rangle,
\label{wf}
\end{equation}
where $T$ denotes time ordering.

Clearly,   the initial condition is $m_0(0)=1/2$.  Expanding the exponential in Eq.~(\ref{wf}) in powers of  the Kondo coupling $\lambda_K$, we find that the first nonvanishing  correction  appears   at  order $J^{2}_{K}$. Following the steps  detailed in Appendix \ref{1stappendix}, we obtain  \be
m_0(\tau)\approx \frac{1}{2}- \frac{\lambda_K^2}{4}\int_{0}^{\tau}d\tau'd\tau''\,C(\tau'-\tau'')+\mathcal{O}(\lambda_K^3),\label{m0C}
\ee
where $C(\tau)=(2\pi\alpha)^{-2}\langle  e^{-i\sqrt{2\pi}\phi_{2s}(0,\tau)} e^{i\sqrt{2\pi}\phi_{2s}(0,0)} \rangle_{0}$   and $\langle \dots \rangle_{0}$ denotes the expectation value in $|\Psi_0\rangle$. 
%The calculation of $m_0(\tau)$ in the case of the impurity coupled to the end of a single wire can be done in a similar fashion. We now apply second-order perturbation theory in the Kondo interaction $H_K^{(2)}$ in Eq. (\ref{HK2}). 
Calculating the correlator and performing the integration over $\tau'$ and $\tau''$, we obtain a  logarithmic dependence for the time decay:\be
m_0(\tau)=\frac{1}{2} - \left( \frac{\lambda_K}{4\pi\alpha\Lambda}\right)^{2}\ln[1+(\Lambda \tau)^{2}]+\mathcal{O}(\lambda_K^3),\label{szt2}
\ee 
where $\Lambda\sim v_F/\alpha$ is an ultraviolet cutoff that appears in the boundary correlators. Note that the leading time dependence of $m_0(\tau)$ stems from the correlator for the operators associated with spin-flip scattering in Eq. (\ref{HK2}), which provides a mechanism for  the relaxation of the polarized impurity spin.

According to the perturbative result  in Eq. (\ref{szt2}), the impurity magnetization would  vanish for times $\tau\sim   \Lambda^{-1}e^{8\pi^2(\alpha\Lambda/\lambda_K)^2}\gg \Lambda^{-1}$. However, we expect the lowest-order result to break down before this condition is reached. The reason is that for times $\tau\sim  \tau_K\sim \Lambda^{-1}e^{\pi v_F/\lambda_K}$ the higher-order corrections must begin to  renormalize the effective Kondo coupling. In fact, the finite time in the quench dynamics plays the role of an effective temperature $T_{\rm eff}\sim 1/\tau$,  which cuts off the infrared singularities associated with  the Kondo effect, see Ref.  \cite{1kondo}. To estimate the Kondo time scale, we set $\Lambda=1$, $v_F=\sqrt2$ for $k_F=\pi/4$, and $\lambda_K=1/2$,  which gives $\tau_K\approx 7\times 10^3$. Thus, in the same way that the Kondo screening cloud can reach mesoscopic  length scales $\xi_K\sim 1\mu$m \cite{affleck2010kondo}, the Kondo time is much longer than the microscopic time scale set by the exchange coupling.  In this work we focus on the  regime $\Lambda^{-1}\lesssim\tau\ll \tau_K$, in which perturbation theory is under control and the dynamics is governed by the ultraviolet (weak coupling) fixed point of the Kondo problem.  This is the  relevant regime for comparison with the numerical results in Sec. \ref{numresults} and possibly with future experiments that might probe the nonequilibrium dynamics in this kind of quench protocol \cite{Kanasz-Nagy2018}.  At long times, $\tau\gg \tau_K$, the dynamics must be governed by the low-energy fixed point at which the impurity spin is completely screened by the conduction electrons \cite{Furusaki1994,Fabrizio1995}.

For the case where the impurity is coupled to both wires, we apply second-order perturbation theory in the Kondo coupling in Eq. (\ref{hk}). We obtain Eq. (\ref{m0C}) with the correlator $C(\tau)$ replaced by $\langle   {F}(\tau) {F}^{\dagger}(0)\rangle_{0}$.    Once again, the spin relaxation is related to the correlator for the operators associated with spin-flip scattering in the Kondo interaction. In fact, the operator $G$ in Eq. (\ref{Gx}) does not appear in this lowest-order result.  % Equation  (\ref{sz}) is general and  valid even in higher dimensions when the appropriate operator $ {F}(\tau)$ is considered. 
Leaving the details of the calculation to Appendix \ref{1stappendix}, here we write down the result:
\begin{widetext}
\begin{equation}
  m_0(\tau)=\frac{1}{2}-2\left( \frac{\lambda_K}{4\pi\alpha\Lambda}\right)^{2} \left( \ln[1+(\Lambda \tau)^{2}] + \frac{2\mc C^2K_{c}^{2}}{1-K_{c}} 
  \left\lbrace 1 - [1+(\Lambda \tau)^{2}]^{\frac{K_{c}-1}{2K_{c}}}\cos\left[\frac{1-K_{c}}{K_{c}}\arctan(\Lambda \tau) \right] \right\rbrace  \right)+\mathcal{O}(\lambda_K^3). \label{szt}
\end{equation}
\end{widetext}
The logarithmic term in this case differs by a factor of 2 from  the corresponding term in Eq. (\ref{szt2}). The decay is also enhanced by the contribution from the tunnelling terms, which depends on Luttinger parameter $K_c$. 
 Note that the exponent $1-K_c^{-1}$ is negative for $K_c<1$, which means that the power law vanishes for $\tau\to\infty$ and  the term $\propto \mc C^2$ converges to a finite value. This happens because the  tunnelling operators in Eq. (\ref{Fx}) have scaling dimension $(1+K_c^{-1})/2$ and are irrelevant at tree level for repulsive interactions \cite{Fabrizio1995,Andergassen2006}. Therefore,  the relaxation of the impurity magnetization    is still predominantly driven by the marginal terms in the Kondo interaction, which give rise to the logarithmic term in Eq. (\ref{szt}).  In the noninteracting limit $K_c\to1$, the tunneling operators become marginal as well, and Eq. (\ref{szt}) reduces to  the pure logarithmic dependence (setting $\mc C=1$):\be
  m_0(\tau)=\frac{1}{2}-4\left( \frac{\lambda_K}{4\pi\alpha\Lambda}\right)^{2}   \ln[1+(\Lambda \tau)^{2}].  \label{purelog}
 \ee
 This is expected because in this case the Kondo interaction in Eq. (\ref{JK1}) can be rewritten in terms of the coupling to a single noninteracting channel , $H_K^{(2)}=2\lambda_K\mb S_0\cdot \psi_+^\dagger(0)(\boldsymbol\sigma/2)\psi^{\phantom\dagger}_+(0)$, where $\psi_+=(\psi_1+\psi_2)/\sqrt2$ annihilates an  electron in the symmetric orbital.

 \begin{table}[t]
  \begin{tabular}{ccccc}
    \hline \hline 
    $U$ & \hspace{0.3cm}  $v_s$ & \hspace{0.3cm} $v_c$ & \hspace{0.3cm} $v_{\rm max}$&\hspace{0.3cm}  $K_c$\\ \hline
    2 & \hspace{0.3cm} 1.09 & \hspace{0.3cm} 1.67& \hspace{0.3cm} 2.06&\hspace{0.3cm} 0.82\\ 
    4 & \hspace{0.3cm} 0.86 & \hspace{0.3cm} 1.81 &\hspace{0.3cm} 2.05&\hspace{0.3cm} 0.71\\ 
    8 & \hspace{0.3cm} 0.56 & \hspace{0.3cm} 1.93 &\hspace{0.3cm} 2.02&\hspace{0.3cm} 0.62\\
    \hline \hline
  \end{tabular}
  \caption{Velocities and Luttinger parameter obtained from the Bethe ansatz solution of the Hubbard model at quarter filling for different values of the interaction $U$. Here $v_{\rm max}$  refers to the maximum velocity calculated from the exact holon dispersion. \label{table_vel}}
  \end{table}

  \section{Numerical Results} \label{numresults}
  
   \begin{figure*}[t]
  \begin{center}
 \includegraphics[width=0.7\linewidth]{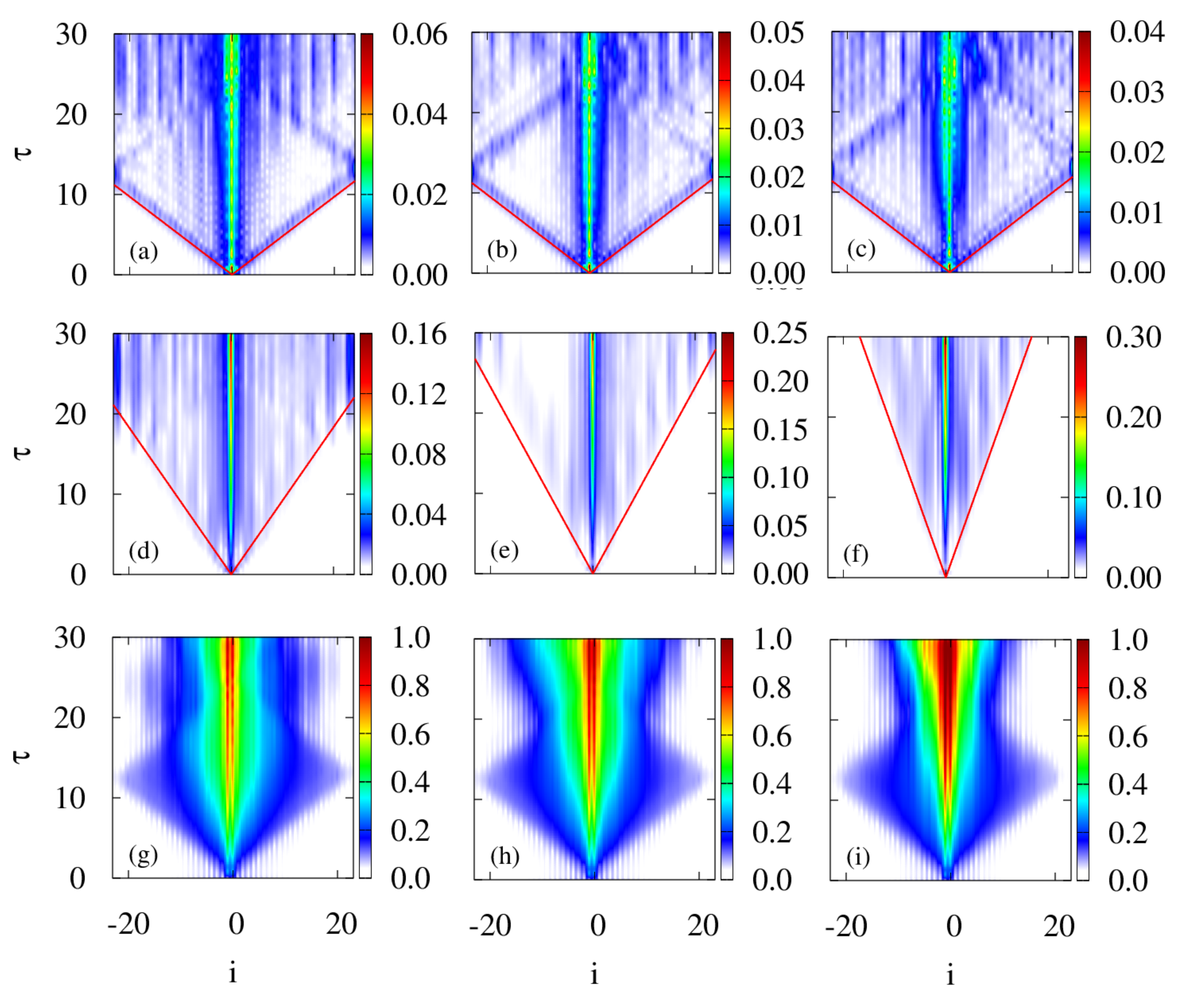}
 \end{center}
 \caption{Spatiotemporal dependence  of local observables and entanglement entropy after the hybridization quench. Here we fix $U_d=7$. Different columns correspond to different  values of the interaction in the Hubbard chains: $U=2$ (left panels), $U=4$ (middle panels), and $U=8$ (right panels). Panels in the same row show the variation in the local density $\Delta \rho(i,\tau)$ [(a)-(c)],    local magnetization $m(i,\tau)$ [(d)-(f)],  and  entanglement entropy $\Delta S(i,\tau)$ [(g)-(i)]. Solid red lines  represent the boundaries of the  light cones defined by the maximum holon velocity $v_{\rm max}$ for $\Delta \rho(i,\tau)$ in the top row and by the spin velocity $v_s$ for $\Delta m(i,\tau)$ in the middle row, as given   in Table~\ref{table_vel}. }
  \label{Figure1}
  \end{figure*}  
  
 In this Section, we use tDMRG to study  the non-equilibrium dynamics following a hybridization quench  described by the Hamiltonian in Eq. (\ref{HamDin}).
 The numerical results  were
obtained  using a second-order Suzuki-Trotter decomposition with a time step   $d\tau =
0.05$, which keeps the error of the order of $10^{-6}$ for the time intervals we consider. 
In the preparation of the initial state, we fix the electron density  at quarter filling, $\rho=0.5$, by adjusting the chemical potential in the chains.  To simulate the dynamics for $\tau>0$, we set the hybridization parameters  $t_{1,2}'=0.5$.  In our DMRG method it is convenient to choose an even number of  sites  for the coupled system. For both geometries, we have fixed the total length to $L=48$ sites. When the impurity is coupled to a single chain,  we use $L_1=0$ and $L_2=47$. When both chains are included, we take  $L_1=23$ and   $L_2=24$.  To lift the Kramers degeneracy in the initial state of the odd-length chain, we apply a weak magnetic field at  the site farthest  from the impurity spin. The small difference between the chains in the second geometry  accounts for a slight asymmetry in the propagation of charge, spin and entanglement in  Fig. \ref{Figure1} below.

  \subsection{Light cones}\label{SecResultsA}

 After the local quench, the propagation of perturbations away from the   impurity site   is bounded by light cones~\cite{LiebRobinson}. It is known that low-lying excitations on top of the ground state can rule  the velocities of the light cones~\cite{Alberto}. For noninteracting leads, $U=0$, the elementary  excitations are electrons that carry both spin and charge simultaneously. By contrast, for $U>0$, the exact solution  of the one-dimensional Hubbard model tells us that  the elementary excitations are spinons, which carry spin, holons and anti-holons, which carry charge, and bound states thereof \cite{essler2005}. We may then anticipate that the light cone velocities  in our quench protocol can be extracted from  the dispersion of spinons and holons in the chains. Importantly, these are bulk properties which do not depend on the parameters  at the impurity site.

We calculate the exact holon and spinon dispersions at quarter filling  by solving the Bethe ansatz integral equations  in the thermodynamic limit following Ref. \cite{essler2005}. Within the Bethe ansatz solution, the ground state is constructed by filling up the spinon and holon states with quantum numbers up to some values, fixed by electron density and magnetization, which define  the Fermi boundary. At zero magnetic field,  the maximum spinon velocity occurs at the Fermi boundary and can be identified with the spin velocity $v_s$ in the TLL theory. On the other hand, the charge velocity $v_c$, calculated from low-energy particle-hole excitations in the holon dispersion, is not the maximum velocity for holons. The reason is that for any density below half filling the inflection point of the holon dispersion lies above the Fermi boundary. Here it is instructive to recall that this is also true for $U=0$, where the free electrons have dispersion $\varepsilon_0(k)=-2\cos k$. The maximum velocity in the band  is $v_{\rm max}=\text{max}\{d\varepsilon_0/dk\}=2$, defined from single-particle states at $k=\pi/2$, which is higher than the Fermi velocity $v_F=2\sin k_F=\sqrt2$ for $k_F=\pi/4$. Remarkably, in the opposite limit $U\to\infty$, the maximum velocity   is also $v_{\rm max}=2$, but here it comes from the dispersion of holons which behave as free spinless fermions  \cite{Ogata1990}.  In Table \ref{table_vel}, we show the values of the spin ($v_s$), charge ($v_c$) and maximum ($v_{\rm max}$) velocities calculated by Bethe ansatz for three different values of $U$. Note that $v_{\rm max}\approx 2$ for all  values  of the interaction.  We stress that   TLL theory does not predict $v_{\rm max}$, as the latter depends on the holon dispersion at finite energies.    
  
In Fig.~\ref{Figure1}, we show the tDMRG results for perturbations in three different quantities as a function of position $i$ and time $\tau$ after the quench when the impurity is coupled to both wires. Let us first discuss the change in the local density, \be
\Delta \rho(i,\tau)=|\left\langle\Psi(\tau) |n_i|\Psi(\tau) \right\rangle -\left\langle\Psi_0|n_i|\Psi_0\right\rangle|,\ee
shown in panels (a)-(c) for $U=2,4,8$, respectively, and fixed $U_d=7$. Since this observable depends on charge fluctuations,  its dynamics  should be dominated by the propagation of holons. The slope of the red lines represented in Figs. \ref{Figure1}(a)-(c) is set  by $v_{\rm max}$ given in Table \ref{table_vel}, in good agreement with the  observed light cones. Note that the fastest holons reach the open boundaries of the finite chains   at time $\tau\approx 12$. After this, a second cone associated with reflection at the boundaries propagates back towards the center of the system. If we are interested in the behavior of local observables in the thermodynamic limit, we must restrict the measurement   to times below this reflection cone.

  Next, we consider the variation in the local magnetization, \be
  \Delta m(i,\tau)=|\left\langle\Psi(\tau) |S^z_i|\Psi(\tau) \right\rangle -\left\langle\Psi_0|S^z_i|\Psi_0\right\rangle|.\ee
  In particular, $\Delta m(0,\tau)=1/2-m_0(\tau)$ corresponds to the decay of the impurity magnetization.  The numerical result is shown in Fig. ~\ref{Figure1}(d)-(f). In this case, the slope of the light cone boundary is defined by the spin velocity $v_s$, which decreases with increasing  interaction. Recall that $v_s\sim 1/U\to 0$ for $U\to\infty$.  Clearly, in the strongly interacting limit  perturbations in the spin sector propagate more slowly than perturbations in the charge sector, {\it cf.} panels (a)-(c).

  \begin{figure}[t]
  \begin{center}
 \includegraphics[width=\columnwidth]{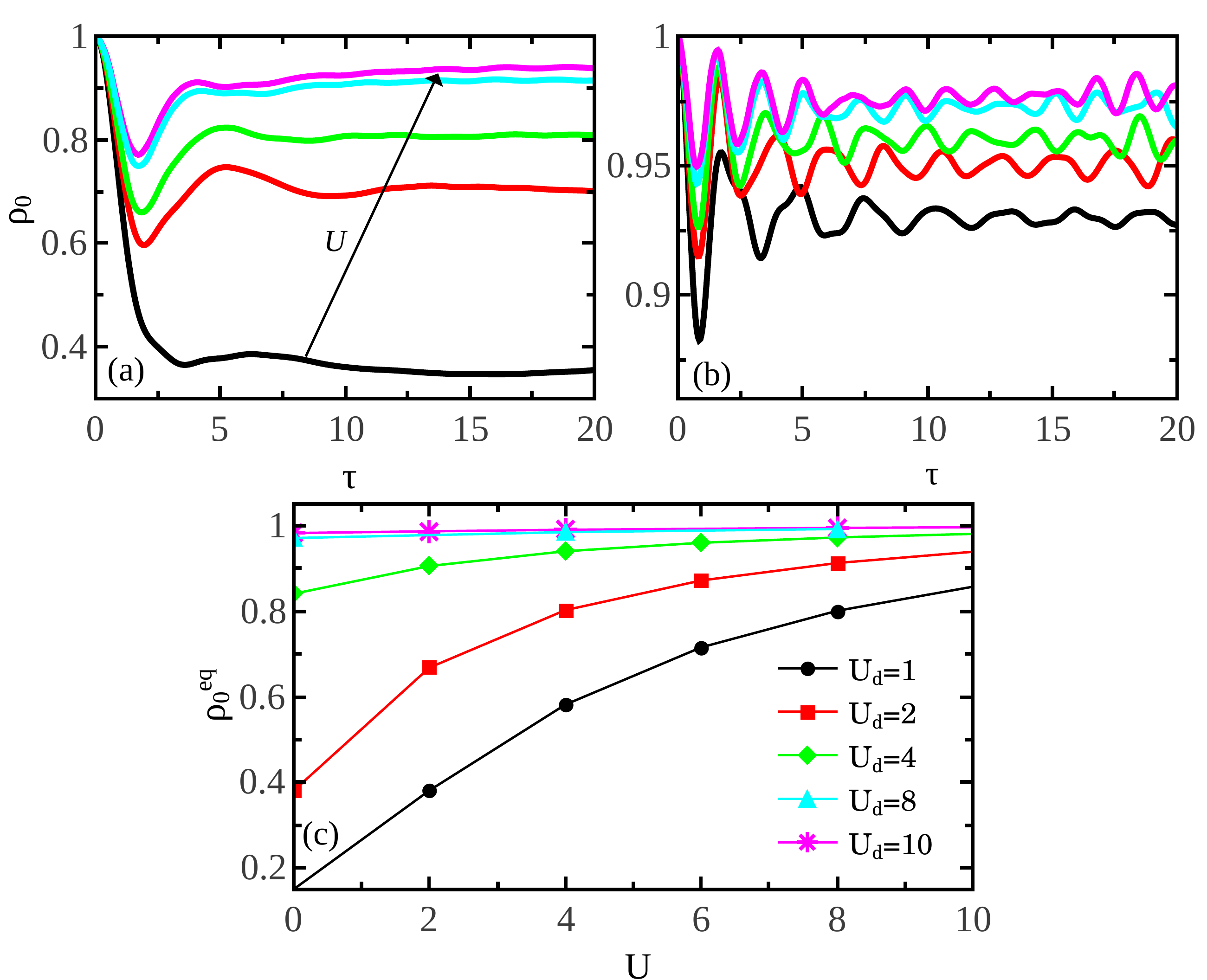} 
 \end{center}
 \caption{Local occupation at the impurity site coupled to two wires. After the hybridization quench, the occupation varies in time as shown in (a) for $U_d=2$ and in (b) for $U_d=7$. Different curves in (a) and (b) correspond to different values of the interaction strength $U$ in the chains; from bottom to top, $U=0,\  2,\ 4, \ 8,\ 10  $.   Panel (c) shows the  equilibrium  occupation as a function of $U$ for different $U_d$, for comparison with the dynamics. }
  \label{Figure3}
  \end{figure}

Finally, we study how the von Neumann entanglement   entropy  (EE) evolves after the quench. Here we divide the system into two partitions, A and B,  such that subsystem A  comprises  the leftmost $i$ sites, while  B contains the $L-i$ rightmost sites. The EE  is defined as  $S(i,\tau)=-\sum_n \xi_n \log \xi_n$, where $\{\xi_n\}$ is the set of eigenvalues of the reduced density matrix $\hat{\rho}_{\rm A/B}(\tau)=\mbox{Tr}_{\rm B/A}|\Psi(\tau)\rangle \langle \Psi(\tau)|$.  In Fig.~\ref{Figure1}(g)-(i), we show the propagation of variations in the EE, given by $\Delta S(i,\tau)=|S(i,\tau)-S(i,0)|$. We observe that the velocity of the   light cone in the entanglement entropy    is  close to   $v_{\rm max}$, indicating that  entanglement propagates with  the velocity of the fastest excitation.

    \begin{figure}[t]
\begin{center}
\includegraphics[width=\linewidth]{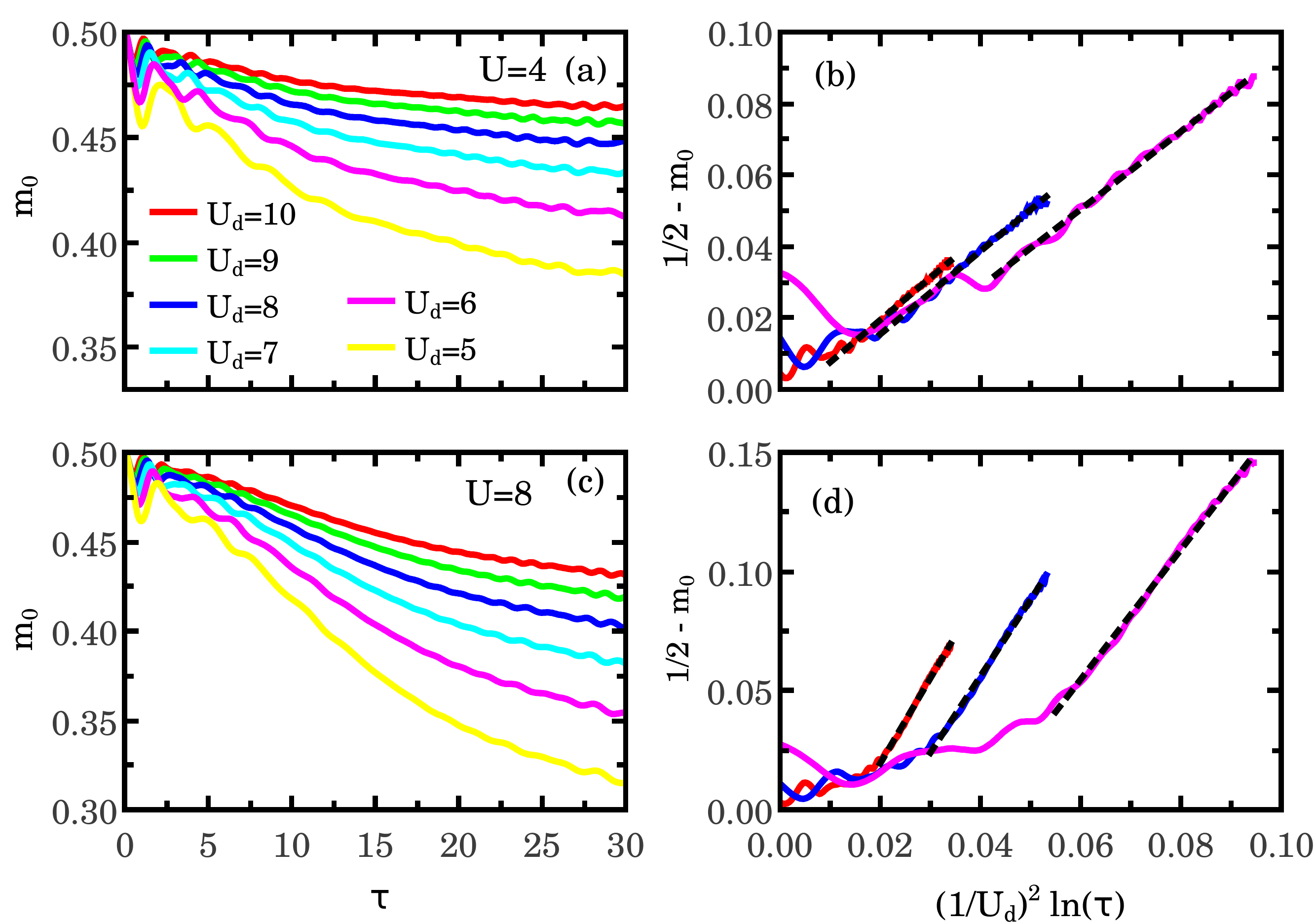}
\caption{Impurity magnetization  as a function of time after the quench  when the impurity spin is coupled to a single chain.  (a) For fixed $U=4$, the magnetization decays more slowly with increasing $U_d$. (b) For times $\tau\gtrsim 10$, the variation  in the magnetization scales logarithmically with time. Panels (c) and (d) show the same as (a) and (b) for $U=8$. The different slopes of the dashed lines in (b) and (d) show that the prefactor of the logarithmic time dependence increases with  $U$. }
\label{Figure4}
\end{center}
\end{figure}

\subsection{Local observables at the impurity site}\label{SecResultsB}

Let us now investigate the time evolution of local observables at the impurity site. Figures~\ref{Figure3}(a) and \ref{Figure3}(b) display the occupation  number $\rho_0(\tau)= \langle \Psi(\tau)|  {n}_0 | \Psi (\tau) \rangle$ for two different values of  $U_d$ and  several  values of  $U$ when the impurity is coupled to both wires.  For comparison, in Fig.~\ref{Figure3}(c) we  plot the corresponding  equilibrium values  $\rho_0^{\rm eq}$ calculated from the ground state of the post-quench  Hamiltonian in Eq.~(\ref{HamDin}). Our results indicate  that   $\rho_0(\tau)$ approaches the equilibrium values for all  values of $U$ and $U_d$,  but it also exhibits  oscillations    with a frequency that increases with   $U_d$.  Most importantly, we observe that the deviation from single occupancy $\Delta \rho_0=1-\rho_0$ decreases as we increase both $U_d$ and $U$, in agreement with the equilibrium results of Ref. \cite{equilibrium}. This implies that the repulsive interaction in the chains facilitates the Kondo regime, where we can neglect charge fluctuations at the impurity site.

    \begin{figure}[t]
\begin{center}
\includegraphics[width=\linewidth]{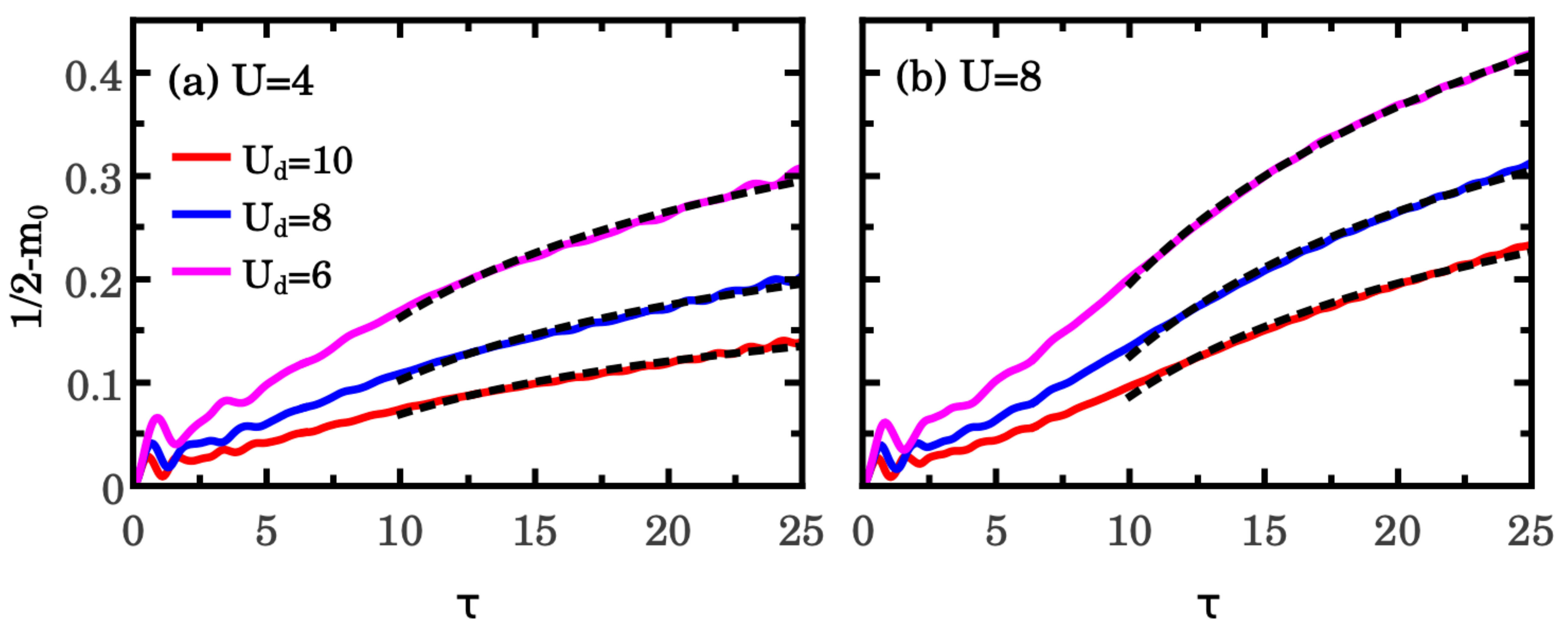}
\caption{Impurity magnetization   when the impurity spin is coupled to both chains.  The results are for (a) $U=4$; (b) $U=8$. The dashed lines   represent the fittings to Eq. (\ref{fit2}) with the parameters given in Table \ref{fitting}. }
\label{Figure5}
\end{center}
\end{figure}

Now we focus on results for $U_d \geq 5$ to ensure that the system is close to the Kondo regime and analyze the decay of the impurity magnetization $m_0(\tau)$. We first discuss the geometry where the impurity is coupled to a single chain, for which we have the simpler analytical expression in Eq. (\ref{szt2}). Figure \ref{Figure4} shows $m_0(\tau)$ for two values of $U$ and several values of $U_d$. The numerical result shows oscillations at short times and also for $\tau\gtrsim 25$. The latter can be attributed  to the finite size effect of reflection at the boundaries. Remarkably, the impurity  magnetization decays faster with increasing $U$ even though the spin velocity $v_s$ decreases with $U$. Note that the total magnetization of the system is conserved due to the SU(2) symmetry of the Hamiltonian, which implies that  $m_0(\tau)$ can only  decay because the magnetization gets transported away from the impurity. However,  the dynamics of the impurity magnetization has a qualitatively different interaction dependence than the ballistic propagation along the light cone. 

At intermediate times, we observe a smooth behavior which can be compared with    the field theory prediction. Approximating Eq. (\ref{szt2}) for $\tau\gg \Lambda^{-1}$ and substituting  $\lambda_K=16(t_2')^2/U_d$, we can write\be
\Delta m(0,\tau)\approx \frac{A}{U_d^2}\ln \tau +\text{const.},\label{approxm01}
\ee
where $A=32(t_2')^4/(\pi\alpha\Lambda)^2$ is a nonuniversal prefactor. In Figs. \ref{Figure4}(b) and \ref{Figure4}(d) we plot $\Delta m(0,\tau)$ versus $U_d^{-2}\ln \tau$. The  results are in agreement  with the logarithmic scaling expected  for $\Lambda^{-1}\ll \tau\ll \tau_K$. Note that  the parameter $A$ (see Table \ref{fitting}) controls the slope of the lines in the semi-log plot. The fact that $A$ remains approximately constant for the larger values of   $U_d$ confirms the dependence of the Kondo coupling $\lambda_K\sim 1/U_d$ for $t_2'\ll U_d$.   On the other hand, $A$ clearly increases with the interaction $U$ in the chains.  This    dependence   is not predicted by TLL theory and must be associated with a renormalization of the cutoff parameters by bulk interactions.  We also note that the logarithmic scaling at intermediate times  demonstrated by our results differs from the exponential decay postulated   in Ref. \cite{Nuss2015}.

\begin{table}[h]
  \begin{tabular}{cccc}
    &   & $U = 4$&  \\
\hline \hline
      $U_d$ & \hspace{0.3cm} $A$ & \hspace{0.3cm} $B$ & \hspace{0.3cm} $C$  \\ \hline
        10 & \hspace{0.3cm} 1.18 $\pm$ 0.03 & \hspace{0.3cm}0.36 $\pm$ 0.07 & \hspace{0.3cm} 0.15 $\pm$ 0.03 \\
        8  & \hspace{0.3cm} 1.17 $\pm$ 0.04 & \hspace{0.3cm} 0.48 $\pm$ 0.02 & \hspace{0.3cm} 0.20 $\pm$ 0.01    \\
        6  & \hspace{0.3cm} 1.10 $\pm$ 0.06 & \hspace{0.3cm} 0.62 $\pm$ 0.04 & \hspace{0.3cm} 0.26 $\pm$ 0.02      \\ 
        \hline \hline  
        \\
            &   & $U = 8$&  \\
\hline \hline
      $U_d$ & \hspace{0.3cm} $A$ & \hspace{0.3cm} $B$ & \hspace{0.3cm} $C$  \\ \hline
        10 & \hspace{0.3cm} 3.57 $\pm$ 0.08 & \hspace{0.3cm}0.71 $\pm$ 0.04 & \hspace{0.3cm} 0.09 $\pm$ 0.01 \\
        8  & \hspace{0.3cm} 3.3 $\pm$ 0.1 & \hspace{0.3cm} 0.80 $\pm$ 0.08 & \hspace{0.3cm} 0.09 $\pm$ 0.01    \\
        6  & \hspace{0.3cm} 2.7 $\pm$ 0.9 & \hspace{0.3cm} 0.79 $\pm$ 0.01 & \hspace{0.3cm} 0.04 $\pm$ 0.01      \\ 
        \hline \hline  
  \end{tabular}
  \caption{Fitting parameters for the decay of the impurity magnetization. The nonuniversal prefactor $A$ is obtained by fitting the tDMRG results for  a single chain to Eq. (\ref{approxm01}). Having fixed $A$, we determine $B$ and $C$ in Eq. (\ref{fit2}) by fitting the data for two chains.  The errors were estimated by varying the time intervals in the fitting. } \label{fitting}
\end{table}

The decay of the impurity magnetization for the  two-wire geometry  is shown in Fig. \ref{Figure5}. In the regime $\tau\gg \Lambda^{-1}$,  the expression in Eq. (\ref{szt})   simplifies to \be
\Delta m(0,\tau)\approx \frac{2A}{U_d^2}\ln\tau -B\tau^{1-K_c^{-1}}+C,\label{fit2}
\ee
where the Luttinger parameter $K_c$ is known from Bethe ansatz (see Table \ref{table_vel}) and $A$, $B$ and $C$ are nonuniversal parameters. Unfortunately, within the limited time range available numerically we are not able to unambiguously distinguish between a pure logarithmic dependence, as expected for the noninteracting case in Eq. (\ref{purelog}), and the combination of a logarithm and a power law with  exponent $1-K_c^{-1}$.  We have fitted the data in Fig.  \ref{Figure5} using Eq. (\ref{fit2}) fixing   $K_c$ from Bethe ansatz and assuming that $A$ takes the same value as in the single-wire geometry, see Fig. \ref{Figure4}.  The parameter $B$ and $C$ were left as fitting parameters. The result of this fitting is represented by the dashed lines in Fig. \ref{Figure5}. The nonuniversal prefactors  are  of order 1 and are  given in Table \ref{fitting}. Overall, the numerical results are consistent with the analytical expressions.

\section{Conclusions\label{sec:conclusions}}

We investigated the role of electronic interactions on the dynamic screening of a localized spin coupled to one-dimensional metallic leads.   We considered a quantum quench in which a magnetic impurity is suddenly connected to interacting chains described by the Hubbard model. We studied  the model numerically via the time-dependent density matrix renormalization group (tDMRG) formalism, as well as analytically within Tomonaga-Luttinger liquid (TLL) theory. Such a theoretical framework gives us access to the evolution of the system at intermediate timescales in regard to the formation of the Kondo  screening cloud.  

We have observed clear signatures of spin-charge separation in the propagation of density and magnetization pulses after the quench.  The propagation is bounded by light cones with velocities that can be extracted from the dispersion relation of the elementary excitations. For densities below half filling, the fastest excitation is a holon with a finite energy above the Fermi level, whose velocity is higher than the charge  velocity used in   TLL theory. This maximum velocity defines the density light cone, while the maximum spinon velocity bounds the magnetization propagation. According to our tDMRG results, the propagation of the entanglement entropy, a nonlocal quantity, is also bounded by the maximum holon velocity.

Concerning local quantities at the impurity site, our results are consistent with relaxation towards the equilibrium values after the local quench. In particular, the  relaxation of the impurity magnetization happens more rapidly if  we decrease the  interaction $U_d$ at the impurity site, thereby enhancing the Kondo coupling, or if we increase the interaction in the chains. While the TLL theory predicts some interaction dependence through the charge Luttinger parameter when the impurity is coupled to two leads, we find that the most important interaction dependence comes from a renormalization of the nonuniversal prefactors. In the case of coupling to a single chain, the field theory  predicts a logarithmic scaling in the perturbative regime of times shorter than the inverse Kondo temperature, which agrees well with our tDMRG results.  

One interesting question that we leave to future work is what happens in the long-time limit, where the spin relaxation must be governed by the low-energy fixed point of the Kondo problem. This regime is hard to access by  numerical techniques such as tDMRG, but one might search for a crossover in the time dependence analogous to the spatial dependence of spin correlations in  the Kondo screening cloud \cite{Eq07,Holzner2009}. It would also be  instructive to consider other geometries, for instance by coupling the impurity site to the middle of a single chain, as in the original studies of the Kondo effect in Tomonaga-Luttinger liquids \cite{Lee1992,Furusaki1994}.

\acknowledgements
We thank Marco Schir\'o for pointing out  his work \cite{Schiro2015}   and for suggesting   that a similar calculation could be done for  a spinful model.
We acknowledge   financial support from CNPq, in particular CNPq INCT-IQ, CAPES, and FAPEMIG. Research at IIP-UFRN is supported by Brazilian ministries MEC and MCTI. 

\appendix
 
%Appendix1

\section{Calculation of the time-dependent impurity magnetization} \label{1stappendix}

In this appendix we provide details of the calculation of $m_0(\tau)$ in the field theory approach. 

Expanding the exponential function in Eq. (\ref{wf}) up to second order in $\lambda_K$, the expectation value of   $ {S}^{z}_{0}$ becomes
\begin{eqnarray}
m_0(\tau) &=&\frac{1}{2} +i\int_{0}^{\tau}d\tau'\langle [ {H}_{K}(\tau'), {S}^{z}_{0}(\tau)] \rangle_{0} \nonumber \\
  && - \frac{1}{2}\int_{0}^{\tau}d\tau'd\tau''\left\lbrace \langle {S}^{z}_{0}(\tau)T[ {H}_{K}(\tau') {H}_{K}(\tau'')] \rangle_{0}
  \right. \nonumber \\
  && \left. + \langle \Tilde{T}[ {H}_{K}(\tau') {H}_{K}(\tau'')] %\times\nonumber\\
   {S}^{z}_{0}(\tau) \rangle_{0} \right. \nonumber\\
  && \left. -2 \langle   {H}_{K}(\tau') {S}^{z}_{0}(\tau) {H}_{K}(\tau'') \rangle_{0} \right\rbrace, 
\end{eqnarray}
where $T$ ($\Tilde{T}$) is the time (anti-time) ordering operator and we omit the lower index $I$ for operators evolved in the interaction picture.

Considering $H_{K}$ given by Eq. (\ref{hk}), it is straightforward to show that
\begin{equation}
\langle [ {H}_{K}(\tau'), {S}^{z}_{0}(\tau)] \rangle_{0}=0,    
\end{equation}
since $[ \mathbf{S}_{0},H_{0}]=0$ for $H_0=\sum_\ell H_\ell^{\rm LL}$. Thus, the first-order term vanishes identically. 

In the second-order terms, it is important to keep track of the  time ordering  in the product of impurity spin operators since\bea
\langle  T  {S}^{+}_{0}(\tau')  {S}^{-}_{0}(\tau'')\rangle_{0}&=&\Theta(\tau'-\tau^{\prime\prime}),\\
\langle  T {S}^{-}_{0}(\tau') {S}^{+}_{0}(\tau'')\rangle_{0}&=&\Theta(\tau^{\prime\prime}-\tau^\prime).
\eea
Similar expressions hold for $\Tilde{T}$. For the two-wire geometry, the $\mc O(\lambda_K^2)$ terms in the  expansion are given by
%\small
\begin{eqnarray}
  & &\langle{S}^{z}_{0}(\tau)T[{H}_{K}(\tau'){H}_{K}(\tau'')] \rangle_{0} \nonumber \\
  & & = \left(\frac{\lambda_K}{2}\right)^{2}\left\lbrace\frac{1}{8}\langle T{G}(\tau'){G}(\tau'')\rangle_{0} \nonumber\right.\\
&&\left. + \frac{\Theta(\tau'-\tau'')}{2}\langle {F}(\tau'){F}^{\dagger}(\tau'')\rangle_{0} \nonumber\right.\\
&&\left. + \frac{\Theta(\tau''-\tau')}{2}\langle {F}(\tau''){F}^{\dagger}(\tau')\rangle_{0}\right\rbrace, 
\end{eqnarray}
%\small
\begin{eqnarray}
  &&\langle \Tilde{T}[{H}_{K}(\tau'){H}_{K}(\tau'')]{S}^{z}_{0}(\tau) \rangle_{0} \nonumber \\
  &&= \left(\frac{\lambda_K}{2}\right)^{2}\left\lbrace \frac{1}{8}\langle \Tilde{T}{G}(\tau'){G}(\tau'')\rangle_{0}\nonumber\right.\\
&&\left.+ \frac{\Theta(\tau'-\tau'')}{2}\langle {F}(\tau''){F}^{\dagger}(\tau')\rangle_{0}\nonumber\right.\\
&&\left.+ \frac{\Theta(\tau''-\tau')}{2}\langle {F}(\tau'){F}^{\dagger}(\tau'')\rangle_{0}\right\rbrace, 
\end{eqnarray}
\begin{eqnarray}
\langle  {H}_{K}(\tau'){S}^{z}_{0}(\tau){H}_{K}(\tau'') \rangle_{0}&=&\left(\frac{\lambda_K}{2}\right)^{2}\left\lbrace \frac{1}{8}\langle {G}(\tau'){G}(\tau'')\rangle_{0}\nonumber\right.\\
&&\left.- \frac{1}{2}\langle {F}(\tau'){F}^{\dagger}(\tau'')\rangle_{0}  \right\rbrace.
\end{eqnarray}
After some algebra, we obtain 
\begin{eqnarray}
m_0(\tau)&=&\frac{1}{2} -\left(\frac{\lambda_K}{2}\right)^{2}\int_{0}^{\tau} d\tau'd\tau''\left\lbrace\langle {F}(\tau'){F}^{\dagger}(\tau'')\rangle_{0}\nonumber\right.\\
&&\left. + \frac{1}{16}\langle [{G}(\tau''),{G}(\tau')]\rangle_{0}\right\rbrace.
\end{eqnarray}
The last term in the  integral vanishes  since the commutator is antisymmetric under the exchange $\tau' \leftrightarrow \tau''$. Therefore, the second-order contribution only involves the correlator for the operator $F(\tau)$.

The correlator can be written in the form 
\be
  \langle {F}(\tau'){F}^{\dagger}(\tau'')\rangle_{0}=\left( \frac{2}{\pi\alpha}\right)^{2}C_{1}(\tau'-\tau'')C_{2}(\tau'-\tau'').\label{FFt}
\ee
The   first factor only involves the spin bosons:
\begin{eqnarray}
C_{1}(\tau)&=&\langle e^{-i\sqrt{\frac{\pi}{2}}\phi_{1s}(\tau)}e^{i\sqrt{\frac{\pi}{2}}\phi_{1s}(0)}\rangle_{0} \nonumber\\
&&\times \langle e^{-i\sqrt{\frac{\pi}{2}}\phi_{2s}(\tau)}e^{i\sqrt{\frac{\pi}{2}}\phi_{2s}(0)}\rangle_{0}\nonumber\\
&=&\frac{1}{1+i\Lambda \tau },
\end{eqnarray}
where we used  $K_{s}=1$ and introduced the high-energy cutoff $\Lambda$. The second factor involves both  charge and spin bosons:\begin{eqnarray}
C_{2}(\tau)&=&\frac{1}{8}\left\lbrace \langle e^{i\sqrt{\frac{\pi}{2}}\phi_{1c}(\tau)}e^{-i\sqrt{\frac{\pi}{2}}\phi_{1c}(0)}\rangle_{0}\times[1\rightarrow 2]\nonumber\right.\\
&&\left. +  \langle e^{i\sqrt{\frac{\pi}{2}}\phi_{1s}(\tau)}e^{-i\sqrt{\frac{\pi}{2}}\phi_{1s}(0)}\rangle_{0}\times[1\rightarrow 2]\right\rbrace\nonumber\\
&=&\frac{1}{8}\left[ \frac{1}{(1+i\Lambda \tau)^{\frac{1}{K_{c}}}} + \frac{1}{1+i\Lambda \tau}\right]. \end{eqnarray}
Finally, the correlator in Eq. (\ref{FFt}) can be written as
\begin{eqnarray}
\langle {F}(\tau'){F}^{\dagger}(\tau'')\rangle_{0}&=&  \frac{1}{2\pi^2\alpha^2}\left\lbrace \frac{1}{[1+i\Lambda(\tau'-\tau'')]^{\frac{1}{K_{c}}+1}}\nonumber \right.\\
&&\left. + \frac{1}{[1+i\Lambda(\tau'-\tau'')]^{2}} \right\rbrace.    
\end{eqnarray}

\bibliographystyle{apsrev4-1}
%\bibliography{references}


\end{document}